\documentclass[preprint,aps,showpacs,nofootinbib]{revtex4}
\usepackage{amsfonts}
\usepackage{graphicx}
\usepackage{epsfig}
\newcommand{\bea}{\begin{eqnarray}}
\newcommand{\eea}{\end{eqnarray}}
\begin{document}
\title{$M_{T2}$-assisted on-shell reconstruction of missing
momenta \\and its application to spin measurement at the LHC}
\author{Won Sang Cho, Kiwoon Choi, Yeong Gyun Kim and Chan Beom Park}

\affiliation{{\it Department of Physics, KAIST, Daejeon 305--017,
Korea} }
\begin{abstract}
We propose a scheme to assign a 4-momentum to each WIMP in new
physics event
producing a pair of mother particles each of which decays to an
invisible weakly interacting massive particle (WIMP) plus some
visible particle(s). The transverse components are given by the
value that determines the event variable $M_{T2}$, while the
longitudinal component is determined  by the on-shell condition on
the mother particle. Although it does not give the true WIMP
momentum in general, this $M_{T2}$-assisted on-shell reconstruction
of missing momenta provides kinematic variables well correlated to
the true WIMP momentum, and thus can be useful for an experimental
determination of new particle properties. We apply this scheme to
some processes
to measure the mother particle spin, and find that spin
determination is possible even without  a good knowledge of the new
particle masses.
\end{abstract}
 \maketitle

The Large Hadron Collider (LHC) at CERN will explore soon the TeV
energy scale where new physics beyond the Standard Model (SM) is
likely to reveal itself. There are two major motivations for new
physics at the TeV scale, one is the hierarchy problem and the other
is the existence of dark matter (DM). Constraints from electroweak
precision measurements and proton decay suggest that TeV scale new
physics preserves a $Z_2$ parity under which the new particles are
odd, while the SM particles are even. Well-known examples include
the weak scale supersymmetry (SUSY) with conserved $R$-parity \cite{susy},
little Higgs model with $T$-parity \cite{littlehiggs}, and extra-dimensional model with
$KK$-parity \cite{extradim}. The lightest new particle in these models is typically
a weakly interacting massive particle (WIMP) which is a good DM
candidate.

A typical LHC signature of new physics model with conserved $Z_2$
parity is the production of a pair of new particles decaying to
invisible WIMPs  plus some visible SM particles. As the WIMP momenta
can not be measured,
an event by event reconstruction of the new particle's rest frame is
not available,  which makes the determination of new particle mass
and spin quite non-trivial. For the mass measurement, several
methods have been proposed so far \cite{mass,lester,masskink}, and
they might work with certain accuracy depending upon the kinematic
situation. One can also attempt to determine the spin  from the
production rate or from a kinematic variable distribution that shows
spin correlation \cite{spin}. However, usually spin determination is
more difficult as it requires a larger statistics and/or a good
knowledge of the mass spectrum and branching ratios.

In this paper, we propose a scheme to assign a 4-momentum to each
WIMP in new physics event, which can provide kinematic variables
useful for the experimental determination of new particle
properties, particularly the spin. We dub this scheme the
''$M_{T2}$-Assisted On-Shell'' (MAOS) reconstruction as the
transverse components are given by the value that determines the
collider variable $M_{T2}$ \cite{lester}, while the longitudinal
component is determined by the on-shell condition imposed on the
mother particle.
This MAOS reconstruction of WIMP momenta can be done even without a
good knowledge of the WIMP and mother particle masses. In the
following, we discuss the MAOS reconstruction in more detail, and
apply it to the 3-body decay of gluino and the 2-body decay of
Drell-Yan produced slepton, as well as to their universal
extra-dimension (UED) equivalents, in order to see the spin effects
in these processes \cite{csaki,barr}.

%
%

To start with, let us consider the new physics process: \bea
\label{new_physics}pp\rightarrow Y(1) +\bar{Y}(2)\rightarrow
V(p_1)\chi(k_1)+V(p_2)\chi(k_2),\eea where $V(p)$ denotes generic
set of visible SM particles with total 4-momentum $p^\mu$ and
$\chi(k)$ is the WIMP with 4-momentum $k^\mu$. For an event set of
this type, one can introduce {\it trial} mother particle and WIMP
masses, $m_Y$ and $m_\chi$, and impose the on-shell condition
together with the missing $E_T$ constraint:
 \bea
\label{onshell} (p_i+k_i)^2=m_Y^2,\quad
  k_i^2=m_\chi^2,
  \quad\mathbf{k}_{1T}+\mathbf{k}_{2T}=\mathbf{p}_T^{\rm miss},
  \eea
where $\mathbf{p}_T^{\rm miss}$ denotes the missing transverse
momentum of the event. As these provide 6 constraints for 8
unknowns, $k_i^\mu$ ($i=1,2$), there are 2-parameter family of
solutions, which can be parameterized by $\mathbf{k}_{1T}$. For any
choice of real $\mathbf{k}_{1T}$, e.g.
$\mathbf{k}_{1T}=\tilde{\mathbf{k}}_{T}$, the longitudinal WIMP
momenta are determined to be \bea \label{twofold} {k}_{iL}=
\frac{1}{(E^V_{iT})^2}\left[ p_{iL}A_i\pm
\sqrt{p_{iL}^2+(E_{iT}^V)^2}\sqrt{A_i^2-(E^V_{iT}E^\chi_{iT})^2}\right]\equiv
\tilde{k}_{iL}^\pm,\eea where
$E^V_{iT}=\sqrt{p_i^2+|\mathbf{p}_{iT}|^2}$,
$E^\chi_{iT}=\sqrt{m_\chi^2+|{\mathbf{k}}_{iT}|^2}$, and
$A_i=\frac{1}{2}(m_Y^2-m_\chi^2-p_i^2)+\mathbf{p}_{iT}\cdot{\mathbf{k}}_{iT}$
for $\mathbf{k}_{1T}=\tilde{\mathbf{k}}_{T}$ and
$\mathbf{k}_{2T}=\mathbf{p}_T^{\rm miss}-\tilde{\mathbf{k}}_{T}$.
It is obvious that $\tilde{k}_{iL}^\pm$ are real {\it if and only
if} \,$|A_i|\geq E^V_{iT}E^\chi_{iT}$ which is equivalent to \bea
\label{realcondition} m_Y \geq \mathrm{max}\{M_{T}^{(1)},
M_{T}^{(2)}\},\eea where
 $M_T^{(i)}=\sqrt{p_i^2 +m_{\chi}^{2}+ 2(\,E^V_{iT}
E^\chi_{iT}-\mathbf{p}_{iT}\cdot\mathbf{k}_{iT})}$ corresponds to
the transverse mass of the mother particle $Y(i)$ with
$\mathbf{k}_{1T}=\tilde{\mathbf{k}}_{T}$ and
$\mathbf{k}_{2T}=\mathbf{p}_T^{\rm miss}-\tilde{\mathbf{k}}_{T}$.


In principle, one could choose event-by-event any value of
$\tilde{\mathbf{k}}_{T}$.  However, the condition
(\ref{realcondition}) suggests that the {\it best} choice  of
$\tilde{\mathbf{k}}_{T}$ is the one minimizing
 $\mathrm{max}\{M_{T}^{(1)}, M_{T}^{(2)}\}$ for each event,
i.e. the value giving the collider variable $M_{T2}$ \cite{lester}:
\begin{equation}
M_{T2}(p_i,m_\chi)\equiv
\min_{\mathbf{k}_{1T}+\mathbf{k}_{2T}=\mathbf{p}_T^{\rm miss}}
\left[ \mathrm{max}\{M_{T}^{(1)}, M_{T}^{(2)}\}\right],
\label{mt2_def}
\end{equation}
where the minimization is performed over all possible WIMP
transverse momenta $\mathbf{k}_{iT}$ under the constraint
$\mathbf{k}_{1T}+\mathbf{k}_{2T}=\mathbf{p}_T^{\rm miss}$. For given
trial masses $m_{\chi,Y}$, this choice of $\tilde{\mathbf{k}}_{T}$,
which is unique for each event, allows the largest event set to have
real $\tilde{k}^\pm_{iL}$. In the following, we call this scheme the
$M_{T2}$-Assisted On-Shell (MAOS) reconstruction, which assigns one
or both of the two 4-momenta \bea
\tilde{k}^{\pm}=(\sqrt{m_\chi^2+|\tilde{\mathbf{k}}_T|^2+|\tilde{k}_L^{\pm}|^2},
\tilde{\mathbf{k}}_T, \tilde{k}_L^{\pm}),\eea to each WIMP in the
new physics event (\ref{new_physics})\footnote{For simplicity, here
we consider the production of $Y\bar{Y}$ in which the up-streaming
momentum is negligible. It is however straightforward to generalize
the MAOS reconstruction to the process with a sizable up-streaming
momentum.}. With these MAOS momenta, one can construct various
kinematic variables whose distribution shape may provide information
on new particle properties.

By construction, if $m_Y\geq M_{T2}^{\rm max}(m_\chi)$, where
$M_{T2}^{\rm max}$ denotes the maximum of $M_{T2}$ over the event
set under consideration, the reconstructed MAOS momenta are real for
all events. However, if $m_Y < M_{T2}^{\rm max}(m_\chi)$,
generically there are some events which do not allow real
longitudinal MAOS momenta. Also by construction, for the $M_{T2}$
end-point events with $M_{T2}=M_{T2}^{\rm max}$, the MAOS momentum
is same as the true WIMP momentum if (i) the trial masses used for
reconstruction are same as the true masses, and (ii) the considered
event set is large enough to give $M_{T2}^{\rm
max}(m_\chi=m_\chi^{\rm true})=m_Y^{\rm true}$. (Note that
$M_{T2}^{\rm max}(m_\chi=m_\chi^{\rm true})\leq m_Y^{\rm true}$ for
generic event (sub)set, and the bound is saturated if the full event
set is taken into account.)
One can also find that $\tilde{k}^+_L=\tilde{k}^-_L$ for an event
whose $M_{T2}$-value is same as $m_Y$, which obviously includes the
end-point event when the trial mother particle mass is chosen as
$m_Y=M_{T2}^{\rm max}(m_\chi)$. In fact, the most interesting
feature of the MAOS momenta, which will be discussed below with a
specific example, is that they are distributed around the true WIMP
momentum with non-trivial correlation even when constructed with
generic trial WIMP and mother particle masses.



\begin{figure}[ht!]
\begin{center}
\epsfig{figure=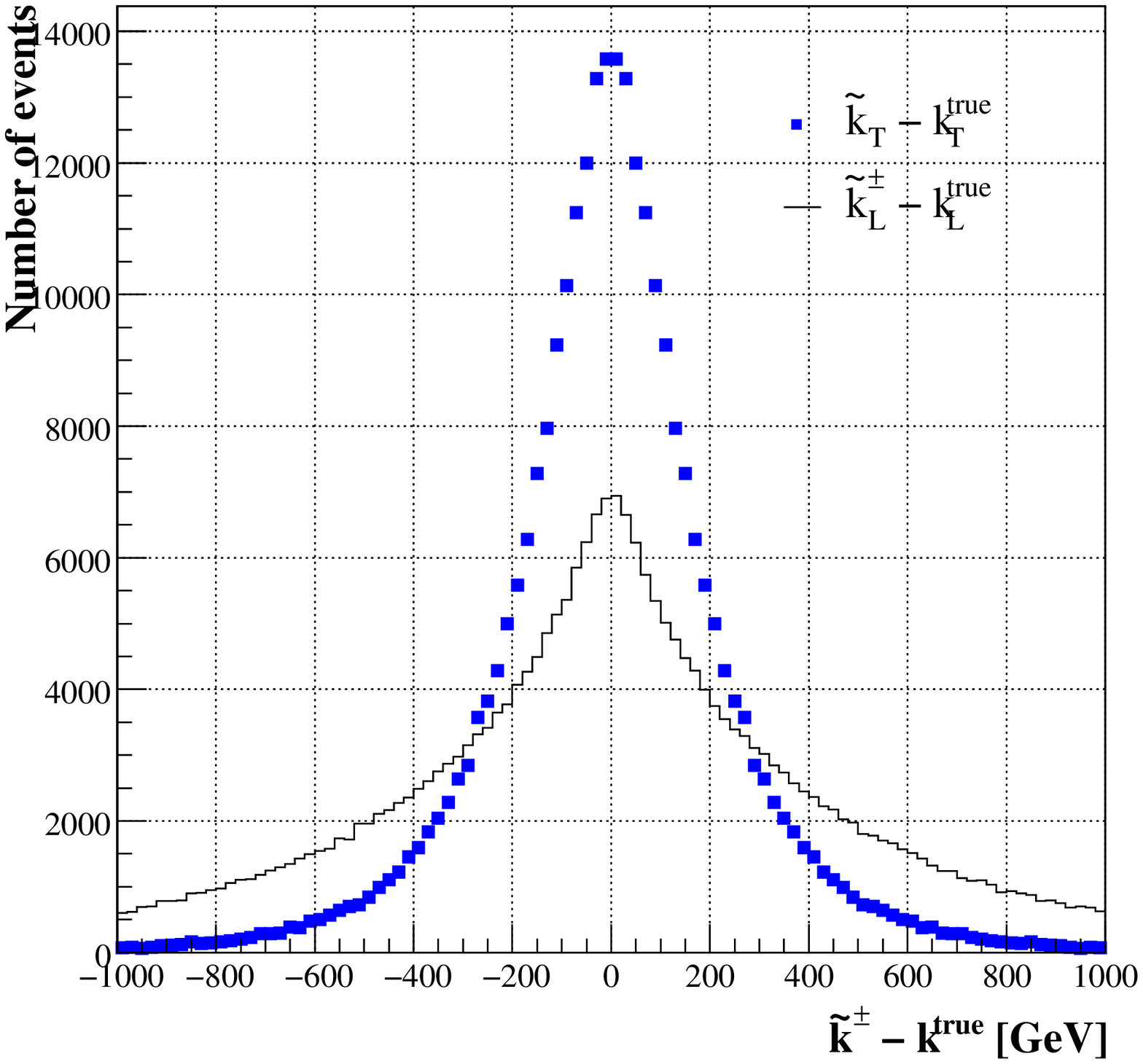,width=6cm,height=6cm,angle=0}
\epsfig{figure=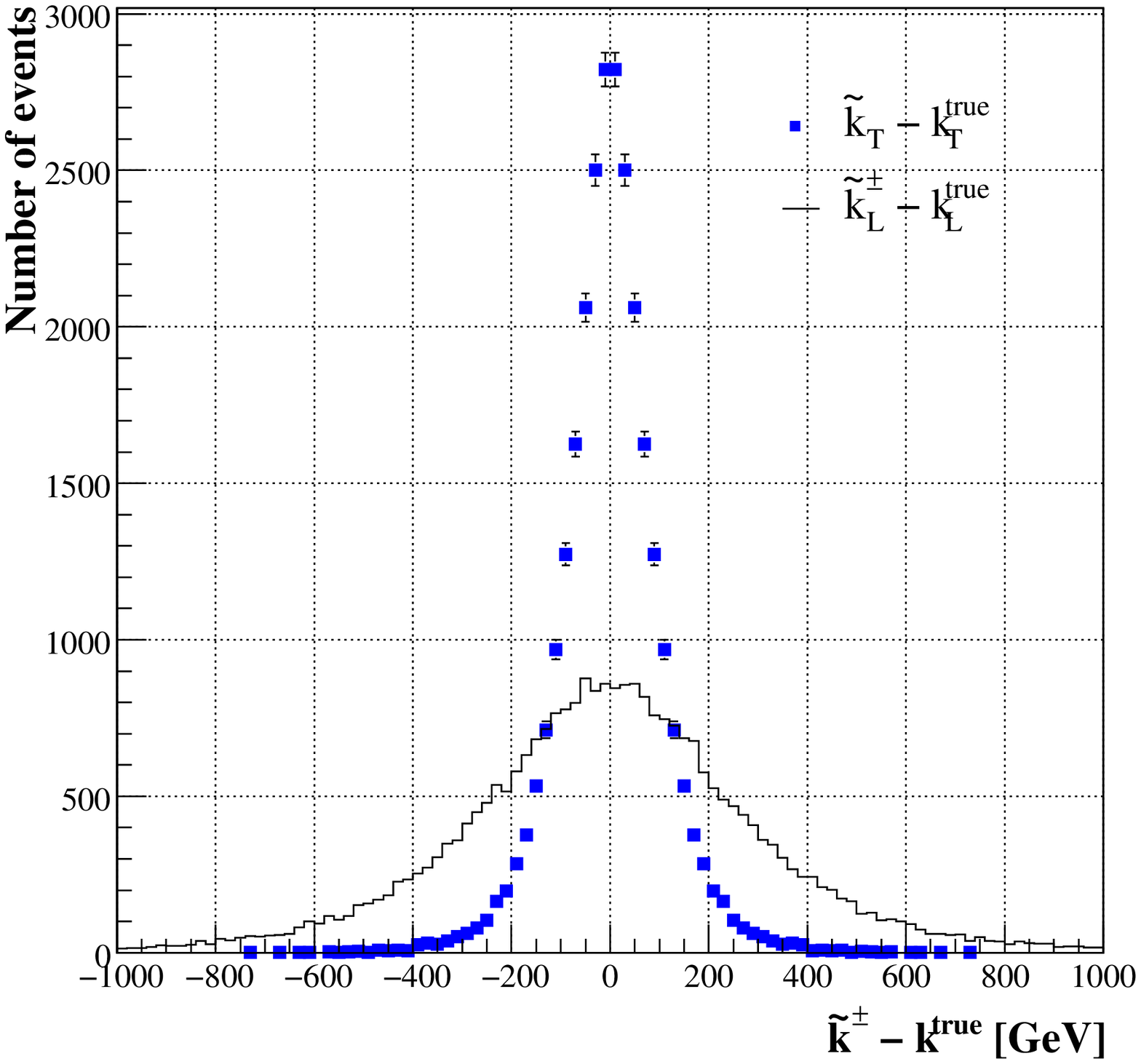,width=6cm,height=6cm,angle=0}
\end{center}
\caption{The distributions of $\tilde{k}^{\pm} - k^{\rm true}$ for
(a) the full event set, and (b) the top 10\% end-point events of
$M_{T2}$.
 Here the MAOS momenta were constructed with $m_{\chi}=0$ and $m_Y=M_{T2}^{\rm max}(m_\chi=0)$.} \label{fig:momentum}
\end{figure}

As a specific application of the MAOS reconstruction, we first
consider the symmetric 3-body decays of gluino pair in SUSY model,
$\tilde{g}+\tilde{g}\rightarrow q\bar{q}\chi +q\bar{q}\chi$, and
also the similar decays of KK gluon pair in UED-like model. For SUSY
case, we choose the focus (SPS2) point of mSUGRA scenario
which gives  the following weak scale masses: $m^{\rm
true}_{\tilde{g}} = 779$ GeV, $m^{\rm true}_\chi = 122$ GeV, and
$m^{\rm true}_{\tilde q} \simeq 1.5$ TeV. We also consider its UED
equivalent in which the gluino is replaced with the first KK gluon
$g_{(1)}$, the Bino $\tilde{B}$ with the first KK $U(1)_Y$ boson
$B_{(1)}$ , and squarks with the first KK quarks. Using {\tt
MadGraph/MadEvent} \cite{mgme}, we have generated the events at
parton-level for both SUSY and UED cases, and constructed the MAOS
momentum $\tilde{k}^\pm$. We then examined the distribution of
$\Delta \tilde{k}\equiv \tilde{k}^\pm-k^{\rm true}$ for both SUSY
and UED event sets with various choices of trial masses $m_{\chi,Y}$
and also of the event subset selected by  their $M_{T2}$ values. We
found that the distribution of $\Delta\tilde{k}_T$ is always peaked
at zero for generic value of $m_\chi$, and its width gets narrower
if one considers an event subset including only the near end-point
events of $M_{T2}$. On the other hand, the distribution of $\Delta
\tilde{k}_L$ is more chaotic, partly because of the error
propagation from $\Delta\tilde{k}_T$ and also the two-fold
degeneracy of the longitudinal component. Still it is peaked at
zero, although the width is significantly broader, for a wide range
of $(m_\chi, m_Y)$ which includes the case with $m_Y= M_{T2}^{\rm
max}(m_\chi)$. As an example,  we depict in Fig. \ref{fig:momentum}
the distributions of $\Delta\tilde{k}_{T,L}$ (including both of
$\tilde{k}_L^\pm$) for the MAOS momenta of the SPS2 SUSY event set,
which has been constructed with $m_\chi=0$ and $m_Y=M_{T2}^{\rm
max}(m_\chi=0)$. Fig. \ref{fig:momentum}a shows the distributions
of the full event set, while  Fig. \ref{fig:momentum}b is for a
subset including
only the top 10\% end-point events of $M_{T2}$. We can see that the
MAOS momentum has a good correlation with the true momentum even for
the full event set, and the correlation becomes  stronger for the
near end-point events of $M_{T2}$. This suggests that if one has an
enough statistics, it can be more efficient to do MAOS
reconstruction using only the near end-point events.

%


%
%

If one could measure all final state momenta  in the 3-body decay,
\bea \label{3body} Y\, \rightarrow \,q(p_q)\bar{q}(p_{\bar
q})\chi(k),\eea where $Y=\tilde{g}$ {or} $g_{(1)}$, and
$\chi=\tilde{B}$ or $B_{(1)}$, the spin of $Y$ can be determined by
the 2-D invariant mass distribution $dN_{\rm
decay}/dm_{qq}^2dm_{q\chi}^2$ for $m_{qq}^2=(p_q+p_{\bar q})^2$ and
$m_{q\chi}^2=(p_q+k^{\rm true})^2$ or $(p_{\bar q}+k^{\rm true})^2$.
However, as the true WIMP momentum  is not available, one could have
only the $m_{qq}^2$-distribution, $dN_{\rm decay}/dm_{qq}^2=\int
dm_{q\chi}^2dN_{\rm decay}/dm_{qq}^2dm_{q\chi}^2$. In \cite{csaki},
it was found that the SUSY and UED $m_{qq}^2$-distributions show a
difference in small $m_{qq}^2$ limit, however this difference might
be difficult to be seen in the real data unless the mass ratio
$m_Y^{\rm true}/m^{\rm true}_\chi$ is quite large, e.g. bigger than
7 or 8.

%
\begin{figure}[ht!]
\begin{center}
\epsfig{figure=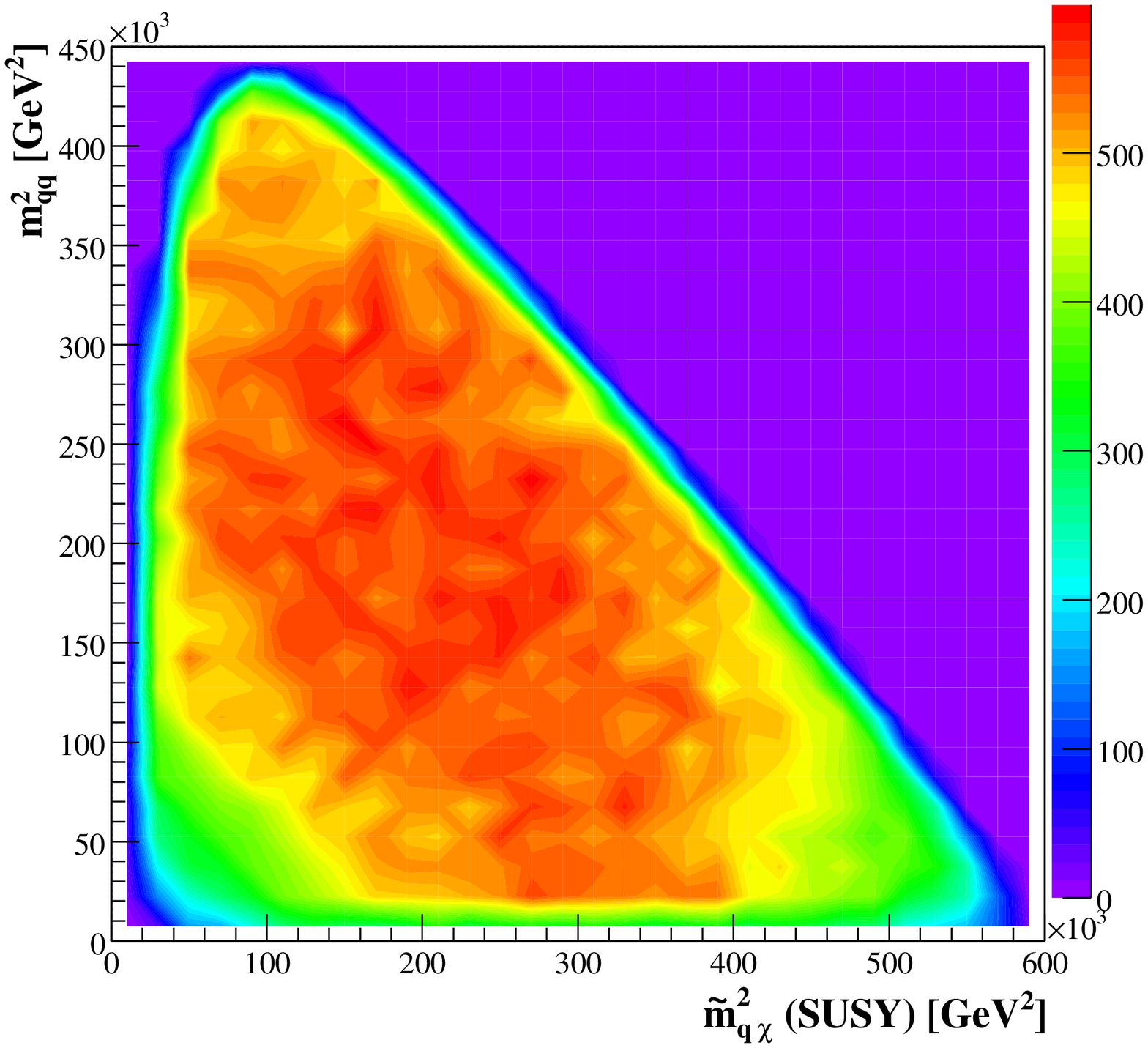,width=6cm,height=6cm,angle=0}
\epsfig{figure=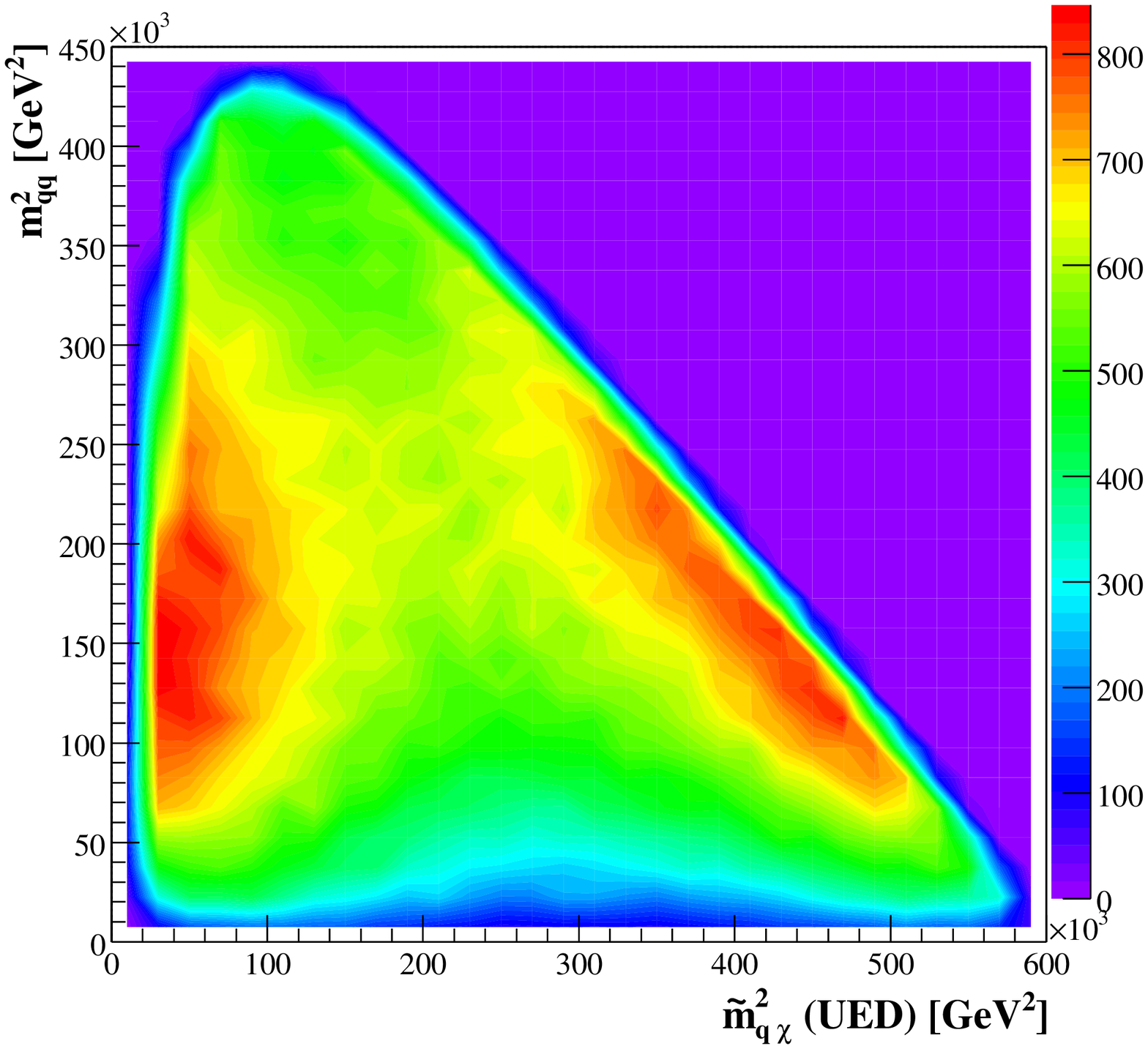,width=6cm,height=6cm,angle=0}
\end{center}
\caption{SUSY and UED Dalitz plots of $m_{qq}^2$ and
$\tilde{m}_{q\chi}^2$ at parton level for
$m_{\chi,Y}=m_{\chi,Y}^{\rm true}$  and very large luminosity.}
\label{fig:dalitz_maos}
\end{figure}

On the other hand, with the MAOS momenta $\tilde{k}^\pm$, we can do
a much better job as the event distribution $dN_{\rm
event}/dm_{qq}^2d\tilde{m}_{q\chi}^2$ is available, where \bea
\tilde{m}_{q\chi}^2= (p_q+\tilde{k}^{\pm})^2 \quad \mbox{or} \quad
(p_{\bar q}+\tilde{k}^{\pm})^2.\eea In Fig. \ref{fig:dalitz_maos},
we depict this MAOS distribution including both $\tilde{k}^+$ and
$\tilde{k}^-$ for the SUSY SPS2 point with
$m_{\chi,Y}=m_{\chi,Y}^{\rm true}$ and its UED equivalent in the
ideal limit of no combinatorial error and very large
luminosity\footnote{In fact, one can use only $\tilde{k}^+$ or only
$\tilde{k}^-$ for the MAOS invariant mass distribution, and still
finds the same shape of distribution.}. The results show a clear
difference, with which one can distinguish SUSY and UED
unambiguously. In fact, these MAOS distributions reproduce
excellently the shape of the true invariant mass distributions
$dN_{\rm decay}/dm_{qq}^2dm_{q\chi}^2$ constructed with the true
WIMP momentum $k^{\rm true}$ \cite{future}.

\begin{figure}[ht!]
\begin{center}
\epsfig{figure=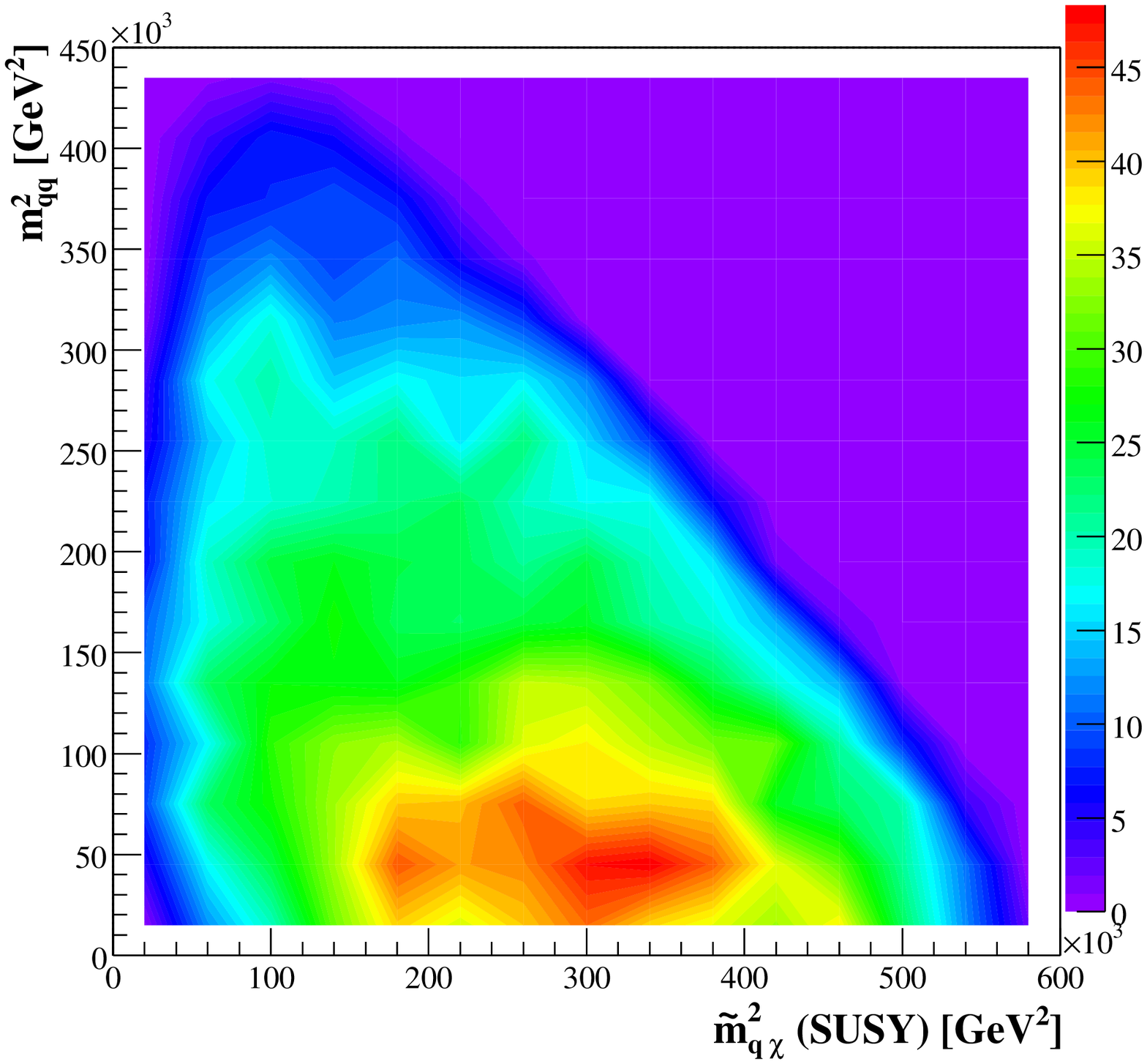,width=6cm,height=6cm,angle=0}
\epsfig{figure=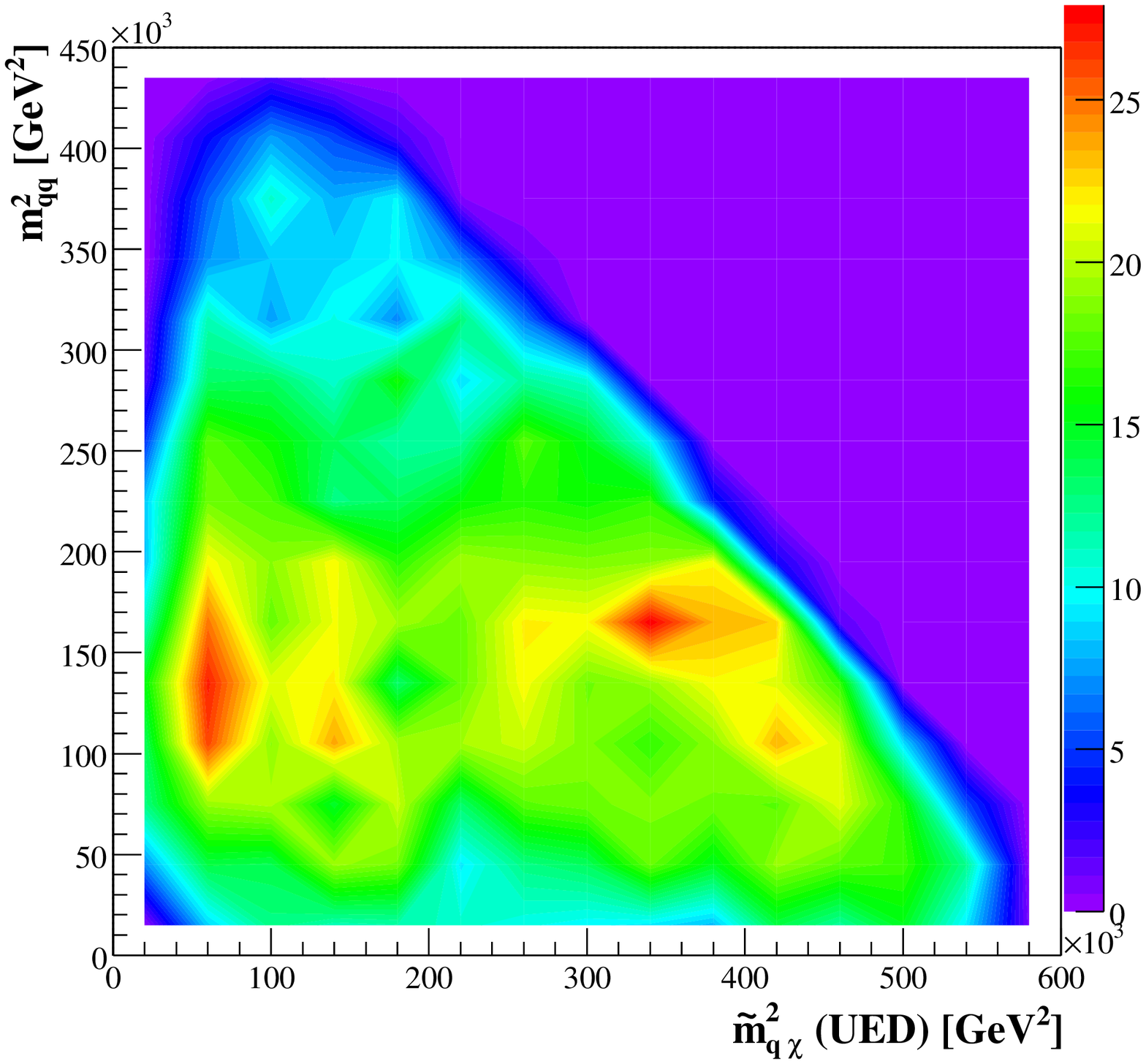,width=6cm,height=6cm,angle=0}
\end{center}
\caption{SUSY and UED Dalitz plots of $m_{qq}^2$ and
$\tilde{m}_{q\chi}^2$ for $m_{\chi,Y}=m^{\rm true}_{\chi,Y}$ and
${\cal L}=300\, {\rm fb}^{-1}$, including the combinatorial error and
smearing effects under a proper event cut.} \label{fig:dalitz_col}
\end{figure}

To see the feasibility of  the MAOS reconstruction in realistic
situation, we analyzed the event sets of the same SUSY and UED
points, but now with the integrated luminosity ${\cal L}=300\, {\rm
fb}^{-1}$. To suppress the backgrounds, we employed an appropriate
selection cut commonly taken for new physics events at the LHC, and
adopted the hemisphere method to deal with the combinatorial
problem. We also included the smearing effects on the momentum
resolution.  The results for $m_{\chi,Y}=m_{\chi,Y}^{\rm true}$ are
depicted in Fig. \ref{fig:dalitz_col}, which still shows
 a clear difference between SUSY and UED.
%

\begin{figure}[ht!]
\begin{center}
\epsfig{figure=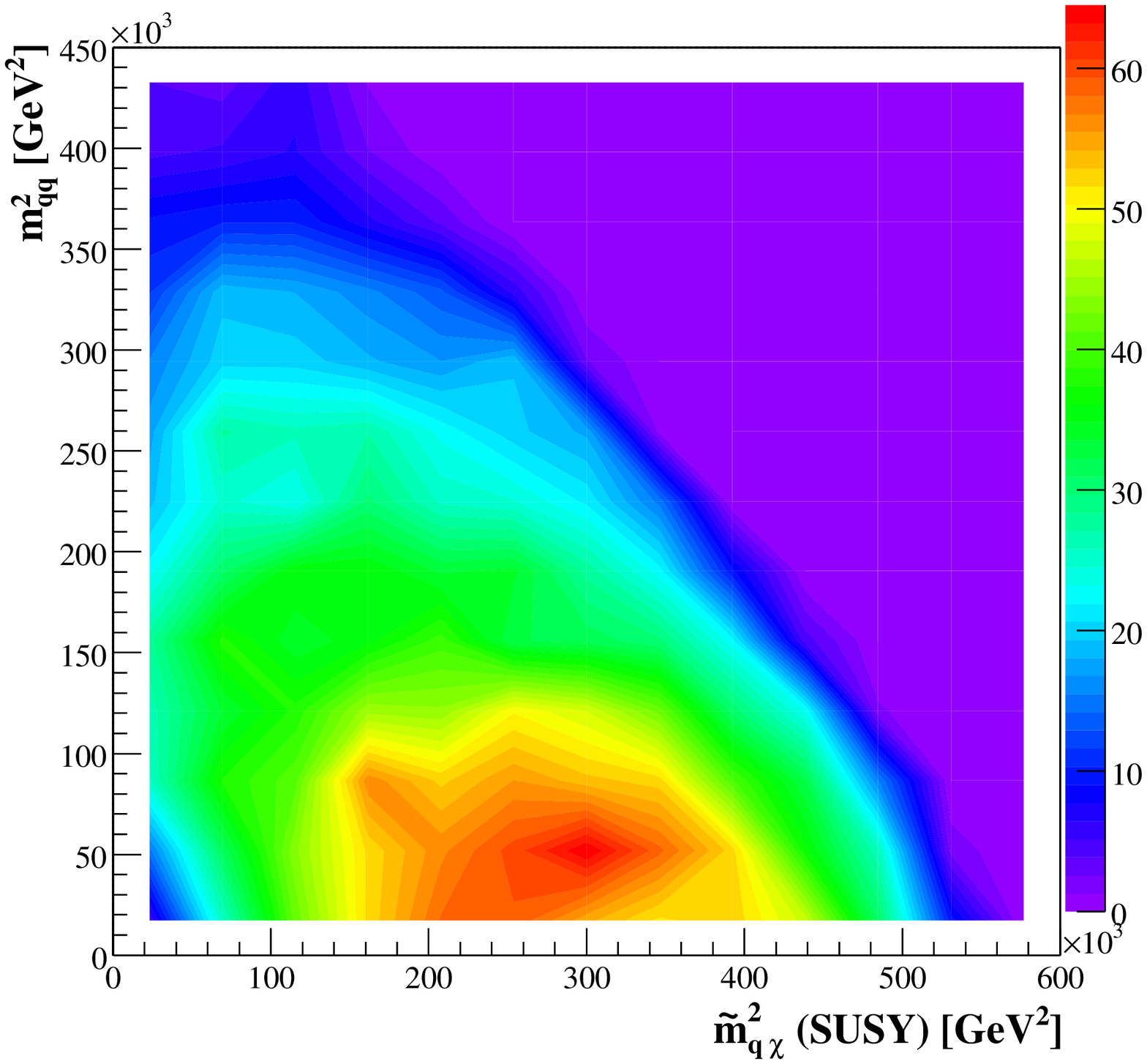,width=6cm,height=6cm,angle=0}
\epsfig{figure=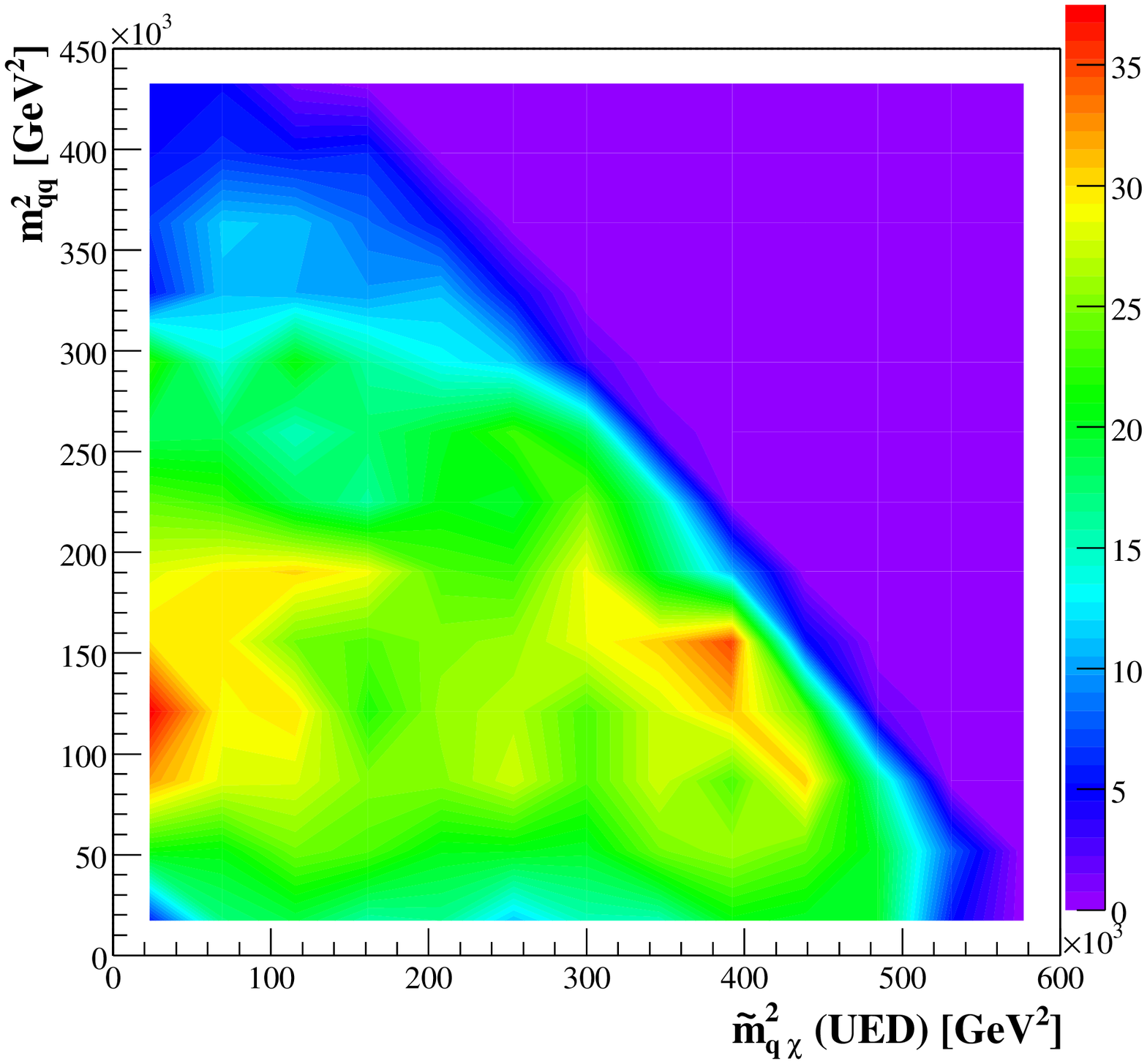,width=6cm,height=6cm,angle=0}
\end{center}
\caption{SUSY and UED Dalitz plots of $m_{qq}^2$ and
$\tilde{m}_{q\chi}^2$ for $m_{\chi}=0$, $m_Y=M_{T2}^{\rm
max}(m_\chi=0)$ and ${\cal L}=300\, {\rm fb}^{-1}$, including the
combinatorial error and smearing effects under a proper event cut.}
\label{fig:dalitz_diff_col}
\end{figure}

A nice feature of the MAOS reconstruction  is that it does not
require any pre-knowledge of the mother particle and WIMP masses. To
see if the spin measurement is possible without  a good knowledge of
new particle masses, we have repeated the analysis for a wide range
of $m_{\chi,Y}$. We then found that the basic feature of
distribution  is retained even when the trial masses used for the
reconstruction are very different from the true masses, and SUSY and
UED can be clearly distinguished in all cases \cite{future}.
We depict the result for one example in Fig.
\ref{fig:dalitz_diff_col}, which is for the case of $m_{\chi}=0$ and
$m_Y=M_{T2}^{\rm max}(m_{\chi}=0)$,  and the result includes the
combinatorial error and detector smearing effects under a proper
event cut.

Once the MAOS WIMP momenta  are available, the MAOS momenta of
mother particles can be reconstructed also.  One can then
investigate the production angular distribution of mother particle
pair in their center of mass frame, which may provide information on
the spin of mother particle. As an example, we have considered the
Drell-Yan pair production of $Y$ ($=$ the slepton $\tilde{l}$ or the
KK lepton $l_{(1)}$) \cite{barr}, \bea q\bar{q}\rightarrow
Z^0/\gamma\rightarrow Y+\bar{Y} \rightarrow l^+(p_1)\chi(k_1)
+l^-(p_2)\chi(k_2),\eea and generated the events for the SPS1a SUSY
point ($m_{\tilde l_R}^{\rm true}=143$ GeV, $m_\chi^{\rm true}=96$
GeV) and its UED equivalent.

As slepton is a scalar particle, the angular distribution is
proportional to $1-\cos^2\theta^*$, where $\theta^*$ is the
production angle with respect to the proton beam direction.  On the
other hand, the corresponding Drell-Yan production of KK leptons
shows the characteristic distribution of spin-half particles, which
is proportional to
$1+\cos^2\theta^*(E_{l_1}^2-m_{l_1}^2)/(E_{l_1}^2+m_{l_1}^2) $
\cite{barr}. One may then examine the lepton angular distribution in
the center of rapidity frame of $l\bar{l}$, which would reflect the
qualitative feature of the above $\theta^*$-dependence of the mother
particle distribution \cite{barr}.

\begin{figure}[ht!]
\begin{center}
\epsfig{figure=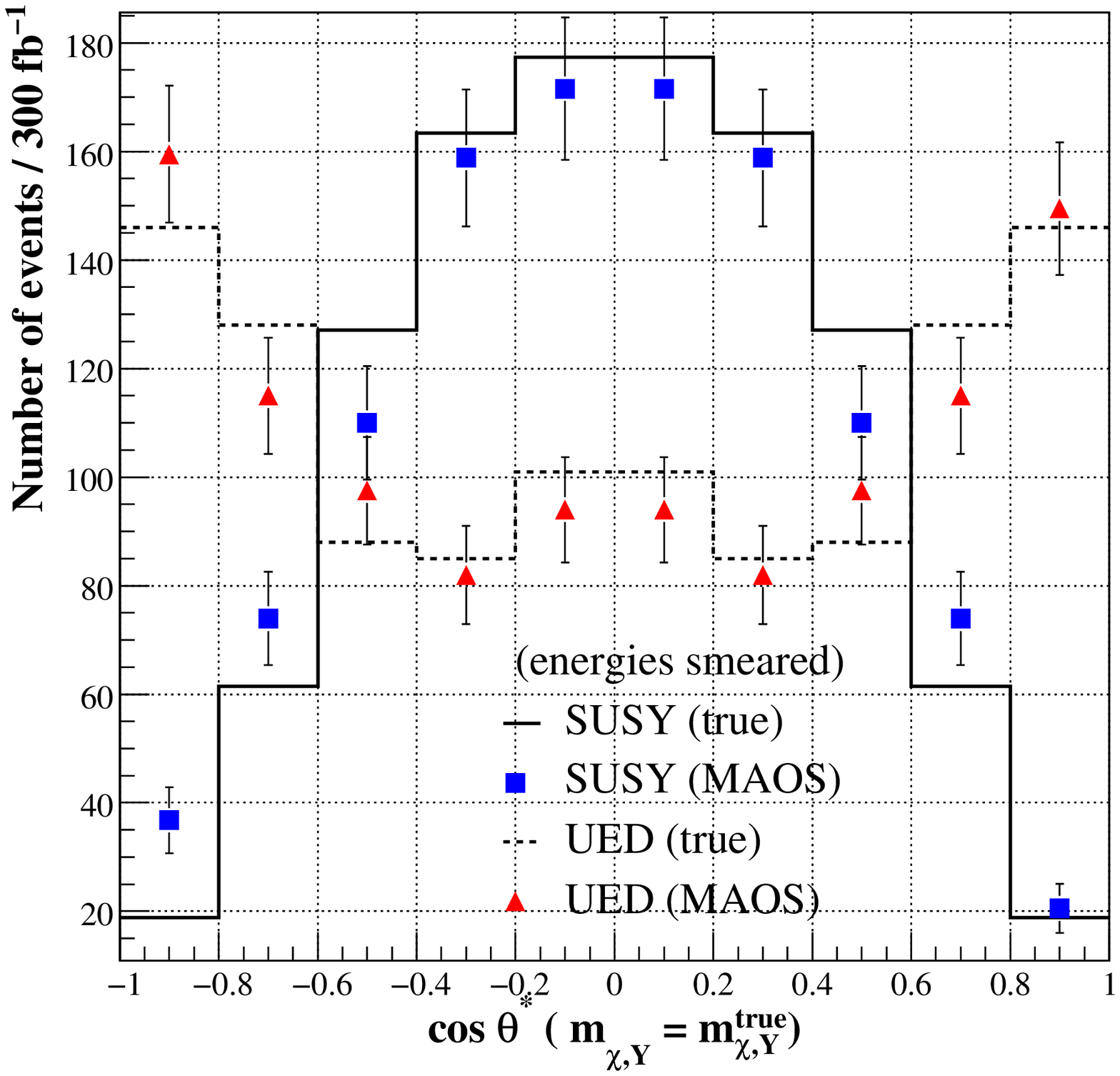,width=6cm,height=6cm,angle=0}
\epsfig{figure=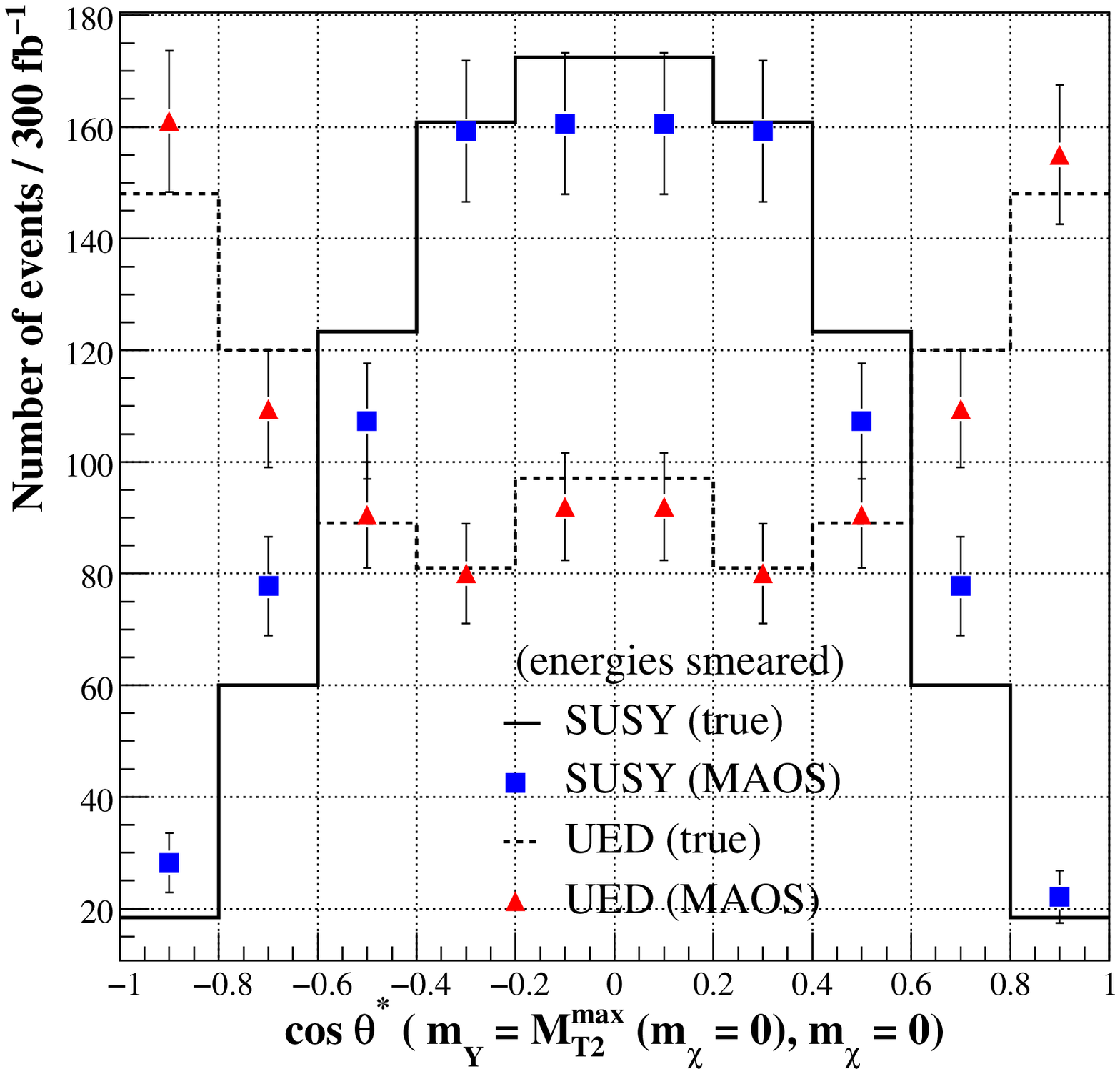,width=6cm,height=6cm,angle=0}
\end{center}
\caption{Slepton and KK-lepton production angular distributions
constructed with the true momenta and the MAOS momenta for ${\cal
L}=300\, {\rm fb}^{-1}$: (a) $m_{\chi,Y}=m_{\chi,Y}^{\rm true}$ and
(b) $m_\chi=0$, $m_Y=M_{T2}^{\rm max}(m_\chi=0)$.}
\label{fig:production_angle}
\end{figure}

Again, we can do a much better job with the MAOS momenta
$\tilde{k}_i^\pm$ ($i=1,2$) as we can probe the angular distribution
of the mother particle MAOS momenta. To see this, we have
reconstructed the MAOS momenta, $p_i+\tilde{k}_i^\pm$, of the
slepton pair and of the KK lepton pair, and examined their angular
distribution in the center of mass frame while including the four
different combinations of MAOS momenta, i.e.
$(\tilde{k}_1^\alpha,\tilde{k}_2^\beta)$ with $\alpha,\beta=\pm$,
altogether. Since it depends on the longitudinal boost to the center
of mass frame, the shape of angular distribution is somewhat
sensitive to the trial mother particle and WIMP masses. To minimize
this sensitivity, we have chosen $m_Y=M_{T2}^{\rm max}(m_\chi)$ and
imposed the event selection cut including only the top 10\% of the
events near the end-point of $M_{T2}$. We also included the detector
smearing effect on the lepton momentum resolution. In Fig.
\ref{fig:production_angle}a, we depict the resulting SUSY and UED
angular distributions for $m_{\chi,Y}=m_{\chi,Y}^{\rm true}$ and
${\cal L}=300\, {\rm fb}^{-1}$, and compare them to the angular
distributions obtained from the true WIMP momenta. The result shows
that the MAOS angular distribution excellently reproduces the true
production angular distribution, with which one can clearly
distinguish SUSY from UED.

To see how much the distribution shape is sensitive to the trial
masses, we repeat the analysis, but now with $m_\chi=0$  and
$m_Y=M_{T2}^{\rm max}(m_\chi=0)$. The result depicted in Fig.
\ref{fig:production_angle}b shows that the distribution has
basically the same shape as the case (Fig.
\ref{fig:production_angle}a) with $m_{\chi,Y}=m_{\chi,Y}^{\rm
true}$.

%
%


To conclude, we have proposed a scheme, the $M_{T2}$-assisted
on-shell (MAOS) reconstruction, to assign
 4-momenta to the WIMP pair in generic new physics event
 of the type (\ref{new_physics}).
Introducing a trial WIMP mass which is common to the whole event set
under consideration, the transverse MAOS momenta of each event are
determined  by the transverse momentum that gives the event variable
$M_{T2}$. On the other hand,
 the longitudinal MAOS momenta are determined by the on-shell
 condition defined with a trial mother particle mass which is again common to the whole event set.
With these MAOS WIMP momenta, one can construct various kinematic
variable distributions which would not be available before, and
extract information on new particle properties from those
distributions. In this paper, we considered an application of the
MAOS reconstruction to the 3-body decays of
 the pair-produced gluinos  and also to the 2-body decays of the Drell-Yan produced
charged slepton pair to discriminate them from their UED
equivalents. Our analysis suggests  that the MAOS reconstruction of
WIMP momenta provides a powerful tool for spin measurement, which
can be viable even when a good knowledge of the new particle masses
is not available.
In the forthcoming paper \cite{future}, we will provide a more
extensive discussion of the MAOS reconstruction and its
applications, including some other processes to measure the new
particle properties.

\begin{acknowledgments}
This work was supported by the KRF Grants funded by the Korean
Government (KRF-2005-210-C0006 and KRF-2007-341-C00010), and the BK21 program of
Ministry of Education.
\end{acknowledgments}


\vfill\eject
\newpage
\end{document}